\documentclass[%
 aip,
jmp,%
 amsmath,amssymb,
preprint,%
]{revtex4-1}

\usepackage{graphicx}
\usepackage{numprint}

\usepackage{dcolumn}
\usepackage{bm}
\usepackage{dcolumn}
\usepackage{amssymb}
\usepackage{booktabs}

\usepackage{color}

\begin{document}


\title{Interfacial energy barrier height of Al$_2$O$_3$/H-terminated (111) diamond heterointerface investigated by X-ray photoelectron spectroscopy}

\author{A. Mar\'echal}
 \email{marechal.aurelien@nims.go.jp}
 \affiliation{National Institute for Materials Science (NIMS), 1-1 Namiki, 305-0044 Tsukuba, Japan}%
\author{Y. Kato}%
\affiliation{ 
National Institute of Advanced Industrial Science and Technology (AIST), 1-1-1 Umezono, 305-8568 Tsukuba, Japan
}%

\author{M. Liao}
\affiliation{%
National Institute for Materials Science (NIMS), 1-1 Namiki, 305-0044 Tsukuba, Japan
}%

\author{S. Koizumi}
\affiliation{%
National Institute for Materials Science (NIMS), 1-1 Namiki, 305-0044 Tsukuba, Japan
}%

\date{\today}

\begin{abstract} 
The interfacial band configuration of the high-$\kappa$ dielectric Al$_2$O$_3$ deposited at 120 $^\circ$C by atomic layer deposition (ALD) on boron- and phosphorus-doped hydrogen-terminated (111) diamond was investigated. Performing X-ray photoelectron spectroscopy measurements of core level binding energies and valence band maxima values, the valence band offsets of both heterojunctions are concluded to be $\Delta E_V$ = 1.8 eV and $\Delta E_V$ = 2.7 eV for the Al$_2$O$_3$/H(111)p and the Al$_2$O$_3$/H(111)n respectively. The ALD Al$_2$O$_3$ bandgap energy was measured from the O 1s photoelectron energy loss spectra to be E$_{G}^{Al_2O_3}$ = 7.1 eV. The interfacial band diagram configuration is concluded to be of type II for both Al$_2$O$_3$/H(111)p and Al$_2$O$_3$/H(111)n heterostructures having conduction band offsets of $\Delta E_C$ = 0.2 eV and $\Delta E_C$ = 1.1 eV respectively. The use of doped (111) hydrogen-terminated diamond for developing future diamond MOSFET is discussed.
\end{abstract}

\maketitle

Roadmaps for future wide bandgap (WBG) semiconductor electronics envision the emergence of SiC and GaN power devices for high voltage applications and energy saving, specifically for applications where high switching speed and low loss devices are required \citep{Okumura2015}. Among WBG semiconductors, Diamond shows the highest calculated figure of merits (FOMs). Thus, it is seen as the ultimate semiconductor for very high power applications owing to its outstanding physical properties. At present, there are still no commercially available diamond power devices, but research is progressing very fast to improve substrate quality, size and device performances. Recent advances have made possible the realization of high performance diamond devices such as Schottky diodes with Baliga's FOM of 75.3 MW.cm$^{-2}$ \citep{Umezawa2013} and 244 MW.cm$^{-2}$ \citep{Traore2014} and high breakdown field larger than 7.7 MV.cm$^{-1}$ \citep{Traore2014}, p-i-n diodes withstanding high reverse bias voltages \citep{Suzuki2013}, junction field effect transistors (JFET) operating at high temperature \citep{Iwasaki2016} and metal-semiconductor field effect transistors (MESFET) with breakdown voltage as high as 1.5 kV \citep{Umezawa2014}. On the other hand, hydrogen-terminated diamond (H-diamond) surfaces have attracted a lot of attention for the development of high performance diamond metal-oxide-semiconductor field-effect transistors (MOSFETs). Indeed, normally-on depletion mode MOSFETs can be fabricated using atomic layer deposition (ALD) of various dielectrics on H-diamond surface \citep{Kawarada2012,Liu2013a,Liu2014}. Also, normally-on or normally-off operation can be achieved depending on the fabrication processes\citep{Liu2015}. High cut-off frequency (50 GHz), high temperature operation (400 $^\circ$C) and high breakdown voltage (600 V) were achieved demonstrating the full potential of diamond MOSFET for power applications\citep{Kawarada2012,Kawarada2014}. These outstanding performances rely on the particular band alignment of the high-$\kappa$ dielectric/H-diamond heterostructure, which is of type II. Such a configuration allows the confinement of a 2D hole layer formed by the elevated barrier height at the high-$\kappa$ dielectric/H-diamond interface. This high barrier height prevents hole leakage to the gate contact when operating the device under accumulation mode (on-state). The off-state is obtained by depleting the channel formed by the 2D hole layer. Another approach is the development of an inversion mode MOSFET, which normally-on operation is obtained by the formation of a minority carrier inversion layer underneath the MOS gate contact in the semiconductor. This structure requires the presence of both electron and hole barrier to prevent gate leakage in accumulation regime as well as in inversion regime. Thus, in order to develop inversion mode diamond MOSFET, ALD Al$_2$O$_3$ was performed on Oxygen-terminated diamond (O-diamond) surfaces and the MOS capacitor electrical properties as well as the interfacial band configuration were investigated \citep{Chicot2013,Marechal2015}. More recently, the operation of a hole-channel inversion mode MOSFET with normally-off operation was reported \citep{Matsumoto2016}. The development of both depletion mode and inversion mode MOSFET is strongly related to two main aspects of the gate dielectric:  i) the quality of its interface with the semiconductor in order to lower the density of interface states and ii) the sufficient band offsets with the semiconductor valence and conduction band affecting the gate leakage currents in the device. The later point have been thoroughly addressed by Liu et al. for various dielectric deposited on H(001)diamond for the development of depletion mode MOSFET. Using X-ray photoelectron spectroscopy (XPS) the interfacial band configuration of several high-$\kappa$ dielectric/H(001) diamond heterostructures were established\citep{Liu2012,Liu2013a,Liu2013b,Liu2016}.
In this work, XPS was used to investigate the interfacial band configuration of the Al$_2$O$_3$/H(111) diamond heterojunction for developing an inversion mode diamond MOSFET. Both boron-doped p-type and phosphorus-doped n-type (111) H-diamond samples were used and the interfacial band diagram of the heterostructures are reconstructed using XPS data. The use of doped (111) H-diamond for the development of future diamond MOSFET is discussed.

\begin{figure}
\includegraphics[width=0.5\textwidth]{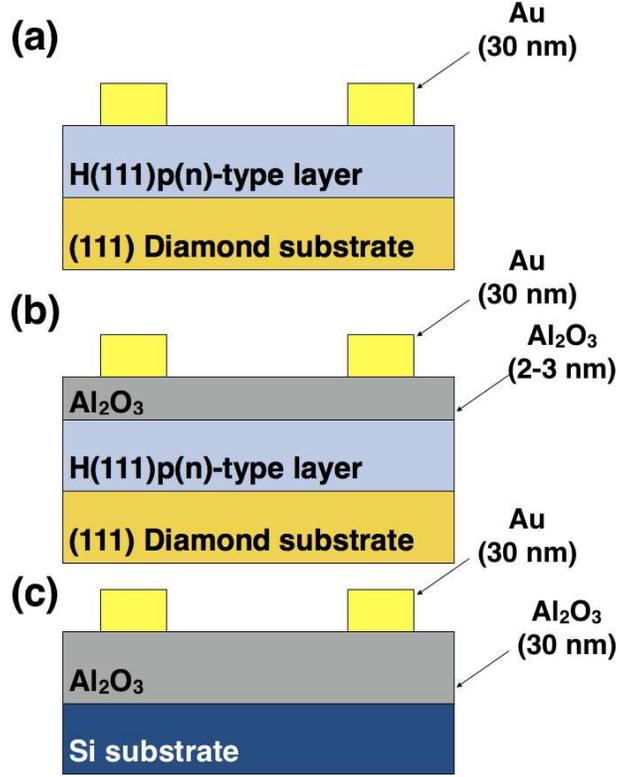}
\caption{\label{fig1}Sample structure used in this study. (a) Hydrogen-terminated p(n)-type (111) diamond used to determine the VBM and the C 1s core level binding energy. (b) Thin Al$_2$O$_3$ layer deposited on H(111)p(n) diamond to determine the Al 2p$_{3/2}$ and C 1s core level binding energies of the heterostructure. (c) 30 thick Al$_2$O$_3$ used to determine the VBM, the Al 2p$_{3/2}$ core level binding energy and the bandgap from the O 1s energy loss spectra.}
\end{figure}

The sample structure used to investigate the interfacial band configuration is shown in Fig.\ref{fig1}. In this study, boron-doped p-type (111) diamond and phosphorus-doped n-type (111) diamond were used (samples labeled H(111)p and H(111)n respectively). All samples were grown using plasma assisted CVD reactors. Details on the growth conditions and reactor can be found elsewhere\citep{Koizumi2000, Koizumi2006, Lazea2012}. Hydrogen termination of the diamond surface was obtained by exposing the diamond surface to a hydrogen plasma at a pressure of 100 Torr at $\sim$900 $^\circ$C for 5 min. Then, Al$_2$O$_3$ was deposited on the hydrogen-terminated (111) diamond surface using ALD at 120 $^\circ$C. The precursor was Trimethylaluminium (TMA) and water was used as oxidant. The thickness of the films were 2 nm and 3 nm on the n-type and p-type (111) diamond respectively (samples labeled Al$_2$O$_3$/H(111)n and Al$_2$O$_3$/H(111)p in the following). These thicknesses are compatible with the calculated inelastic mean free path of ca. 25 \AA \ for C 1s photoelectrons (E$_K$ $\approx$ 1200 eV) in Al$_2$O$_3$ according to TPP-2M equation \citep{Tanuma2005}. These samples enabled to measure the C 1s and the Al 2p$_{3/2}$ core levels. On the other hand, H(111)p(n) samples were used to measure the C 1s core level and the valence band maximum (VBM). The bulk properties of Al$_2$O$_3$ were measured on a 30 nm thick ALD Al$_2$O$_3$ layer deposited on an Si substrate using the same condition as for the Al$_2$O$_3$/H(111) diamond samples. Since the Al$_2$O$_3$ layer deposited by ALD at low temperature is amorphous \citep{Pinero2014}, it is assumed that the layer properties will not be affected by the substrate. Consequently, this sample was used to measure the Al 2p$_{3/2}$ core level, the VBM and the Al$_2$O$_3$ bandgap. Finally, in order to monitor the charge-up effect during the XPS measurement process, gold (Au) pads with thickness of 30 nm and diameter of 300 $\mu$m were deposited on each samples using electron beam evaporation at room temperature.

The XPS measurements were carried out in a Kratos AXIS Nova system at room temperature under base pressure of $\sim$10$^{-9}$ Torr. The samples were illuminated by a monochromated Al K$\alpha$ source (1486.6 eV) with an incident angle of 45$^\circ$ and a collection angle of 90$^\circ$, both relative to the sample surface. The core level spectra were recorded with an energy step of 0.1 eV and at a constant pass energy of 20 eV with an energy resolution of ca. 0.5 eV. Since the X-ray beam area is about 3 mm$^2$, the slot aperture was selected to 110 $\mu$m in order to acquire the signal from a limited area on the sample surface, i.e. to acquire the signal coming from either Au pads or diamond surface. Conductive carbon tape was used to stick the samples on the XPS sample holder. A second piece of carbon tape enabled to contact the samples surface with the sample holder. As described above, Au pads were used to monitor the charge-up effect on the samples. Tab.\ref{tab1} shows the Au 4f$_{7/2}$ binding energies and full width at half maximum (FWHM) measured on each samples. It can be noticed that no difference in the binding energy position is observed from sample to sample. Also, the  Au 4f$_{7/2}$ peak binding energy is in good agreement with the values found in database \citep{nist}. Moreover, the FWHMs are found in close agreement with reported data \citep{Kono2012} and considering the reference value of 84.0 eV for the Au 4f$_{7/2}$ peak position the charge-up effect is assumed to have a negligible influence on the energy shift.
\begin{table}
\caption{\label{tab1} XPS peak energy positions and FWHM of Au 4f$_{7/2}$ core level measured on H(111)n, Al$_2$O$_3$/H(111)n, H(111)p, Al$_2$O$_3$/H(111)p and Al$_2$O$_3$/Si respectively.}
\begin{ruledtabular}
\begin{tabular}{cccccc}
 &\multicolumn{2}{c}{Au 4f$_{7/2}$}\\
 Sample& peak binding energy (eV) & FWHM (eV) \\ \hline
 H(111)n & 84.1 & 0.6\\ \hline
 Al$_2$O$_3$/H(111)n & 84.1 & 0.6 \\ \hline
 H(111)p & 84.1 & 0.6 \\ \hline
 Al$_2$O$_3$/H(111)p  & 84.1 & 0.6 \\ \hline
 Al$_2$O$_3$/Si & 84.1 & 0.7\\
\end{tabular}
\end{ruledtabular}
\end{table}

\begin{figure}
\includegraphics[width=0.85\textwidth]{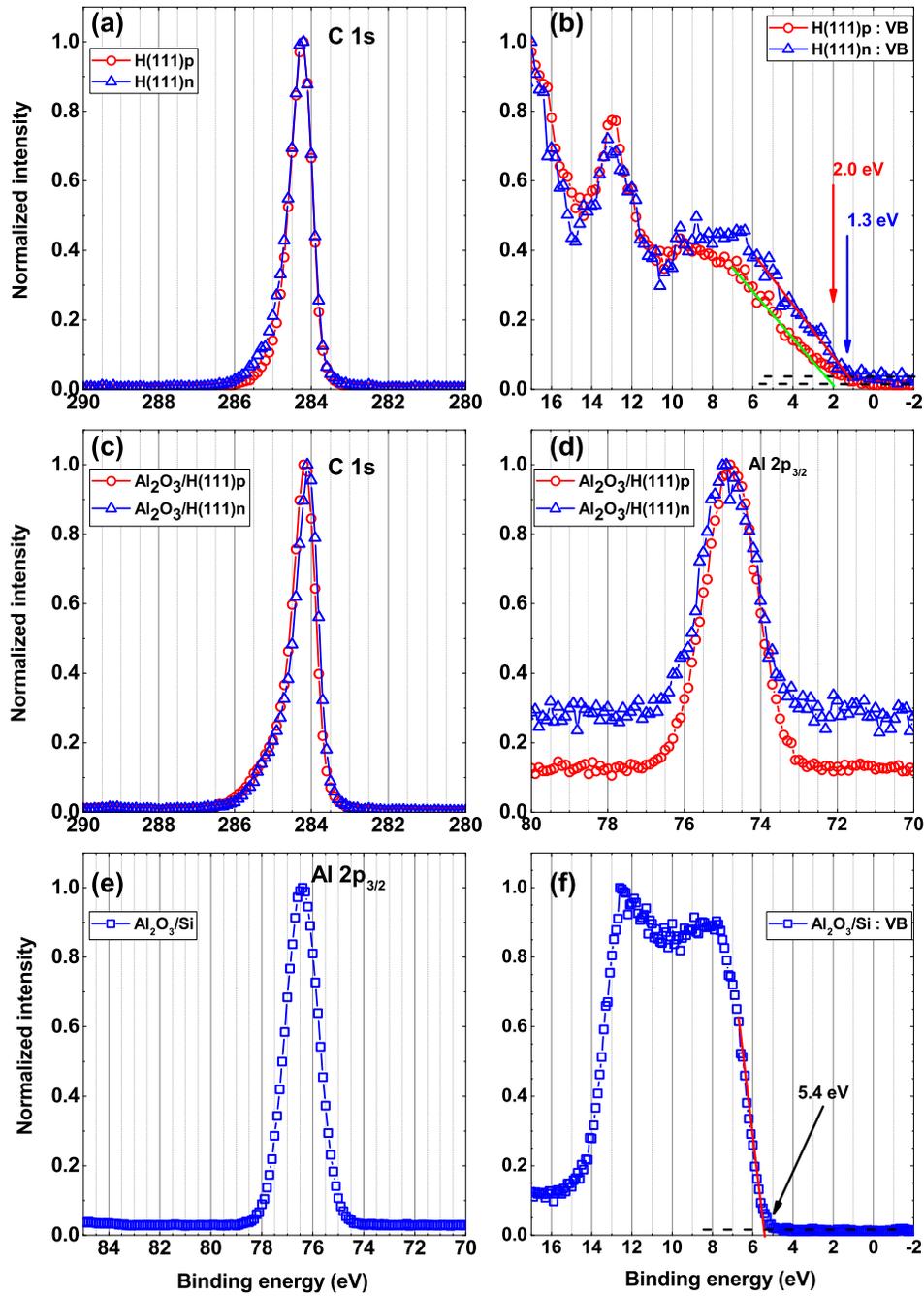}
\caption{\label{XPSSpectra} (a) C 1s and (b) Valence band (VB) XPS spectra from samples H(111)p and H(111)n. (c) C 1s (d) and Al 2p$_{3/2}$ XPS spectra of samples Al$_2$O$_3$/H(111)p and Al$_2$O$_3$/H(111)n. (e) Al 2p$_{3/2}$ and (f) Valence band (VB) XPS spectra of sample Al$_2$O$_3$/Si. Intensities are normalized to unity.}
\end{figure}

The determination of the valence band discontinuity at the Oxide/Semiconductor interface relies on the accurate measurement of the valence band maximum (VBM) with respect to the core levels. This is performed using Kraut et al. \citep{Kraut1980} approach in which the VBM of each material is measured relative to a core level (reference energy) and the core levels are compared in the heterostructure, consisting of the thin Al$_2$O$_3$ layer deposited on H(111) diamond in the present work. Fig.\ref{XPSSpectra} shows the C 1s and valence band (VB) XPS spectra of the H(111)p(n) samples [Fig.\ref{XPSSpectra}.(a) and Fig.\ref{XPSSpectra}.(b)], the C 1s and Al 2p$_{3/2}$ XPS spectra of the Al$_2$O$_3$/H(111)p(n) samples [Fig.\ref{XPSSpectra}.(c) and Fig.\ref{XPSSpectra}.(d)] and the Al 2p$_{3/2}$ and VB XPS spectra of the bulk ALD Al$_2$O$_3$ deposited on Si [Fig.\ref{XPSSpectra}.(e) and Fig.\ref{XPSSpectra}.(f)] . On the core level spectra the peak binding energy positions were extracted using Voigt function, the convolution of a Lorentzian function (related to carrier lifetime broadening) with a Gaussian function (due to the finite resolution of the analyzer) whereas the background was subtracted using Tougaard function. Three components were used to fit the C 1s photoelectron spectra. However, only the main peak contribution is of interest since it is attributed to the C$-$C bond. On the other hand, only one contribution was used to fit the Al 2p$_{3/2}$ implying that no Al$-$C bonds exist at the interface, consistent with previous reports \citep{Liu2012}. The VBM was extracted as the intersection point of a linear regression line covering the leading edge of the valence band photoemission spectra with the background as described by Chambers et al. \citep{Chambers2004}. Tab.\ref{tab2} summarizes the C 1s and Al 2p$_{3/2}$ binding energies, the FWHM of the main peak contributions and the VBM.

\begin{table*}
\caption{\label{tab2} XPS peak energy positions, FWHMs and VBMs of C 1s and Al 2p XPS peaks measured on H(111)n, Al$_2$O$_3$/H(111)n, H(111)p, Al$_2$O$_3$/H(111)p and Al$_2$O$_3$/Si respectively. VBM stands for valence band maximum measured on bulk materials.}
\begin{ruledtabular}
\begin{tabular}{cccccc}
 &\multicolumn{2}{c}{C 1s}&\multicolumn{2}{c}{Al 2p} & VBM\\
 Sample& peak binding energy & FWHM & peak binding energy & FWHM &  \\
 & (eV) & (eV) & (eV) & (eV) & (eV) \\ \hline
 H(111)n & 284.2 & 0.6 & \textemdash & \textemdash & 1.3\\ \hline
 Al$_2$O$_3$/H(111)n & 284.1 & 0.6 & 74.9 &1.6 & \textemdash\\ \hline
 H(111)p & 284.2 & 0.5 & \textemdash & \textemdash & 2.0\\ \hline
 Al$_2$O$_3$/H(111)p  & 284.2 & 0.7 & 74.8 & 1.6 & \textemdash\\ \hline
 Al$_2$O$_3$/Si & \textemdash & \textemdash & 76.4 & 1.5 & 5.4
\end{tabular}
\end{ruledtabular}
\end{table*}

\begin{figure}
\includegraphics[width=0.5\textwidth]{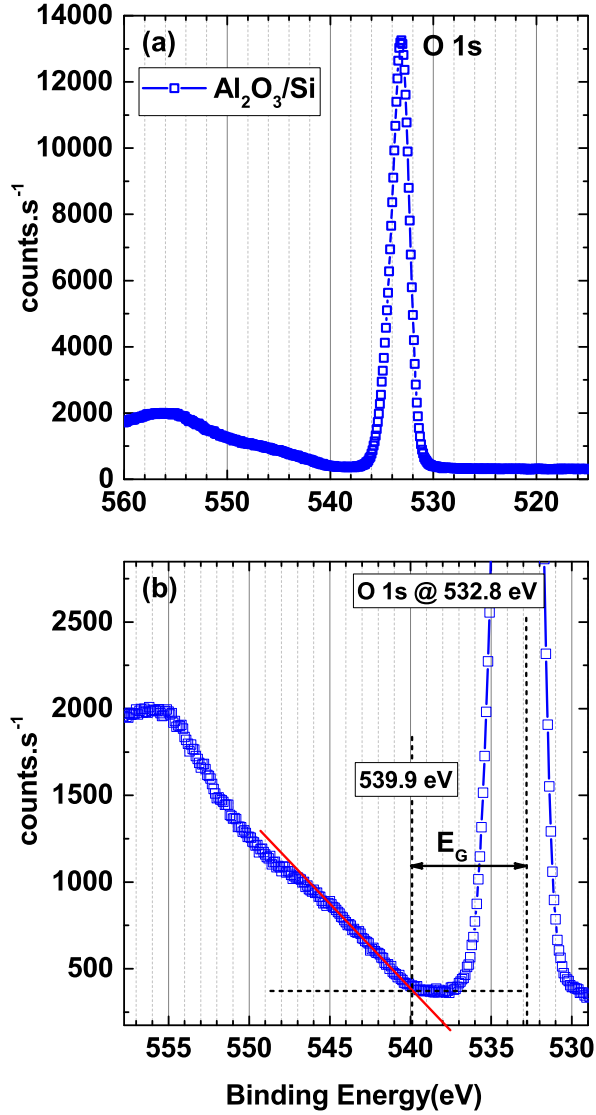}
\caption{\label{O1s} (a) O 1s spectrum of the 30 nm thick Al$_2$O$_3$ layer deposited on Si. (b) Electron energy loss spectrum emphasizing the Al$_2$O$_3$ bandgap of $E_G^{Al_2O_3}$ = 7.1 eV.}
\end{figure}

The valence band offsets are calculated using the following formula\citep{Kraut1980}:
\small
\begin{eqnarray}\label{eq1}
\Delta E_V &=& \left(E_{C1s} - E_{VBM}\right)_{H(111)p(n)} - \left(E_{Al2p} - E_{VBM}\right)_{Al_2O_3/Si} \nonumber\\ 
&-& \left(E_{C1s} - E_{Al2p}\right)_{Al_2O_3/H(111)p(n)},
\end{eqnarray}
\normalsize
where $\left(E_{C1s} - E_{VBM}\right)_{H(111)p(n)}$ is the difference in binding energy between the C 1s core level and the valence band maximum of the hydrogen-terminated (111) diamond, $\left(E_{Al2p} - E_{VBM}\right)_{Al_2O_3}$ is the difference in binding energy of the Al 2p$_{3/2}$ core level and the VBM of the 30 nm thick Al$_2$O$_3$ layer and $\left(E_{C1s} - E_{Al2p}\right)_{Al_2O_3/H(111)p(n)}$ is the difference in binding energy between the C 1s core level and the Al 2p$_{3/2}$ core level of the Al$_2$O$_3$/H(111)p(n) diamond heterostructure. Inserting the binding energies of Tab.\ref{tab2} in Eq.\ref{eq1}, the valance band offsets of the Al$_2$O$_3$/H(111)p and Al$_2$O$_3$/H(111)n heterojunctions are calculated to be $\Delta$E$_V$ = 1.8 eV and $\Delta$E$_V$ = 2.7 eV respectively.
Moreover, the determination of the conduction band offsets requires the measurement of the Al$_2$O$_3$ bandgap which was performed by measuring the O 1s energy loss spectra of the thick Al$_2$O$_3$ sample. The procedure consists in the determination of the intersection point between a linear fit to the O 1s photoelectron energy loss spectra close to the onset of inelastic scattering and the background level. Thus, the bandgap energy is given by the energy difference between the O 1s binding energy and the intersection point energy \citep{Nichols2014}. Fig.\ref{O1s} represents the O 1s photoelectron energy loss spectra and highlights the value of the Al$_2$O$_3$ bandgap of E$_{G}^{Al_2O_3}$ = 7.1 eV. This value is in close agreement with previously reported data for low temperature ALD Al$_2$O$_3$ bandgap of 7.2 eV \citep{Liu2012}. As mentioned above, the Al$_2$O$_3$ layer might be amorphous and the influence of the substrate on the growth quality is negligible as it is evidenced when comparing the Al 2p$_{3/2}$ binding energies and VBM measured on H(001) diamond reported by Liu et al. \citep{Liu2012} with the values presented in Tab.\ref{tab2}. However, the ALD growth conditions such as growth temperature, TMA exposure time, water exposure time and purge time influence the quality of the ALD layer \citep{Groner2004}. 

\begin{figure}
\includegraphics[width=0.6\textwidth]{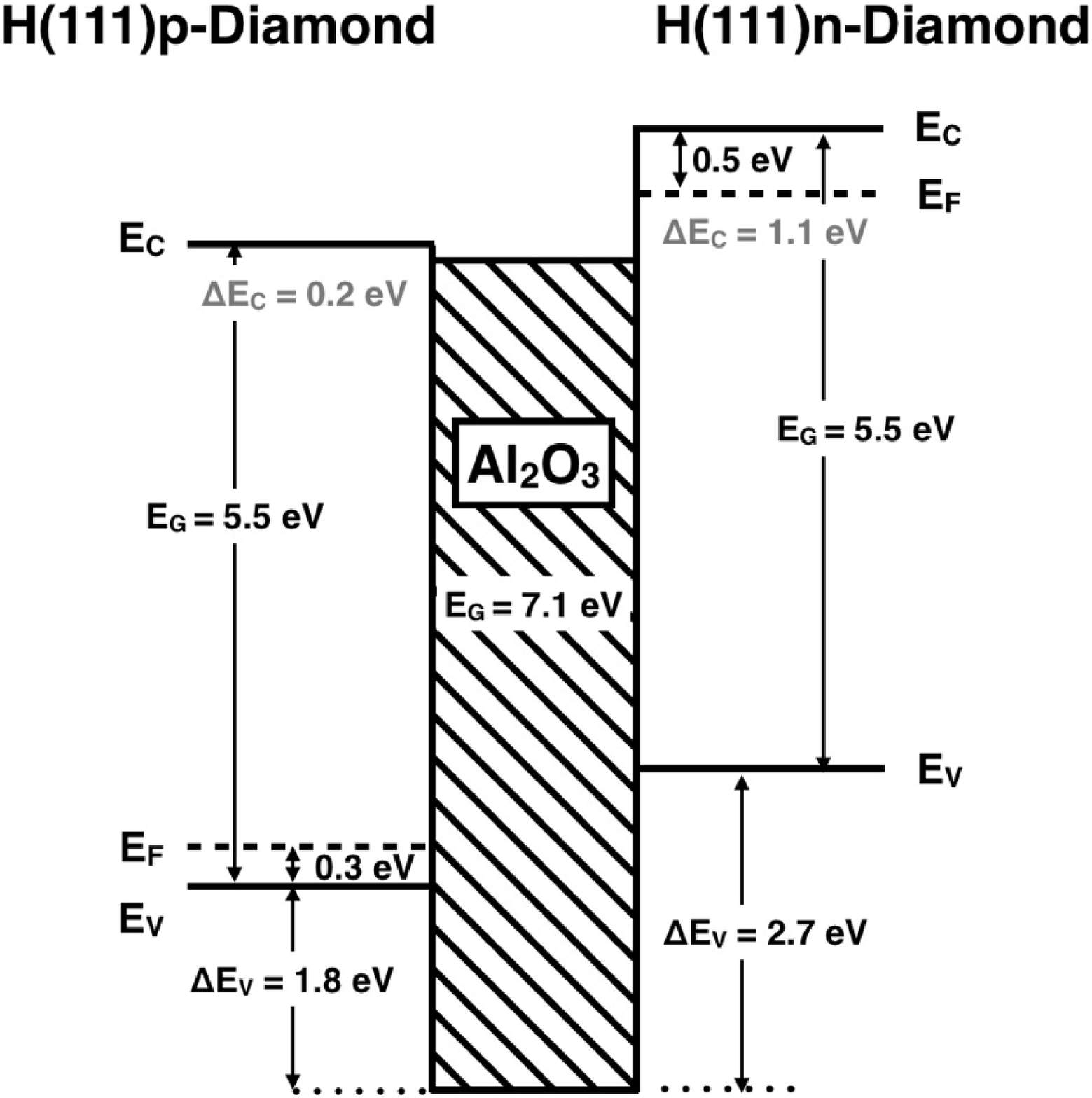}
\caption{\label{flatband} Type II band diagram configuration showing the valence band offset deduced from Eq.\ref{eq1} and the conduction band offset deduced after measurement of the Al$_2$O$_3$ bandgap using Eq.\ref{eq2}. The Fermi level position $E_F$ is calculated under thermal equilibrium in the Boltzmann approximation \citep{Sze2007} taking into account the incomplete ionization of dopants for a doping level $N_A$=$N_D$=10$^{17}$ cm$^{-3}$ with 5$\%$ compensation.}
\end{figure}

The conduction band offsets $\Delta E_C$ are determined using the following formula:

\begin{equation}\label{eq2}
\left|\Delta E_C\right| = E_G^{Al_2O_3} - E_G^{diamond} - \Delta E_V,
\end{equation}

with $E_G^{Al_2O_3}$ = 7.1 eV is the ALD Al$_2$O$_3$ bandgap, $E_G^{diamond}$ = 5.5 eV is the diamond bandgap and $\Delta E_V$ is the valence band offset. As a consequence, values of $\Delta E_C$ = 0.2 eV and $\Delta E_C$ = 1.1 eV are deduced for the Al$_2$O$_3$/H(111)p and the Al$_2$O$_3$/H(111)n heterojunctions respectively. With the knowledge of bandgaps and band offsets values, the band diagram presented in Fig.\ref{flatband} is drawn. Both Al$_2$O$_3$/H(111)p and Al$_2$O$_3$/H(111)n heterostructures show a type II band alignment in agreement with interfacial band configuration reported for high-$\kappa$ dielectric (001) H-terminated diamond heterojunctions\citep{Liu2012,Liu2013a,Liu2013b,Liu2016}. On the other hand a type I band alignment can be found for either CaF$_2$/H-terminated (001) diamond \citep{Liu2013c} or Al$_2$O$_3$/O-terminated (001) diamond \citep{Marechal2015} heterojunctions. 

In the present work, both Al$_2$O$_3$/H(111)p and Al$_2$O$_3$/H(111)n heterojunctions show energy barrier heights for holes, i.e. valence band offsets which could be suitable to perform a MOS gate contact with low gate leakage current. However, in the perspective of developing inversion mode MOSFETs, large energy barrier heights for minority carriers are necessary. This means that the Al$_2$O$_3$/H(111)p MOS structure would not be appropriate for electron inversion channel since there is no barrier to avoid their leakage from the channel to the gate contact. On the other hand, considering the Al$_2$O$_3$/H(111)n MOS structure in which the inversion layer consists of holes, an energy barrier height of $\Delta E_V$ = 2.7 eV would prevent their leakage to the gate contact. Also, if a normally-off operation is achievable, i.e. if the upward band bending of the hydrogen-terminated n-type (111) diamond surface can be preserved \citep{Kono2007}, the absence of barrier for electrons would not be an issue. Indeed, the accumulation regime is not necessary for operating such a device. Thus it is expected that ALD Al$_2$O$_3$ deposited on hydrogen-terminated phosphorus-doped (111) diamond would be suitable to perform a gate contact for next generation diamond power MOSFET.

In summary, interfacial energy barrier heights of ALD Al$_2$O$_3$ deposited on phosphorus-doped and boron-doped hydrogen-terminated (111) diamond were investigated using X-ray photoelectron spectroscopy. The valence band offsets of both Al$_2$O$_3$/H(111)n and Al$_2$O$_3$/H(111)p as well as the ALD Al$_2$O$_3$ bandgap were measured in order to reconstruct the energy band diagrams. It is concluded that the band diagrams are of type II for both heterojunctions similarly to reported results observed on H(001) diamond \citep{Liu2012}. Due to the large barrier height of  $\Delta E_V$ = 2.7 eV measured on the Al$_2$O$_3$/H(111)n heterostructure, this structure is expected to be suitable for developing inversion mode p-channel diamond power MOSFET.

\begin{acknowledgments}
One of the authors (A.M.) would like to thank Dr. J. W. Liu for valuable discussions. This work was supported by the Japan Society for the Promotion of Science (JSPS). 
\end{acknowledgments}

\end{document}